\documentclass{article}%
\usepackage{amsmath}
\usepackage{amsfonts}
\usepackage{amssymb}
\usepackage{graphicx}%
\providecommand{\U}[1]{\protect\rule{.1in}{.1in}}
\begin{document}

\title{Quantum Information as a Measure of Multipartite Correlation}
\author{\textit{Simon J.D. Phoenix}\\Khalifa University, PO Box 127788, Abu Dhabi, UAE}
\maketitle

\begin{abstract}
The mutual information has been successfully used as a measure of correlation
strength between quantum systems, especially for bipartite systems. Here we
examine the use of this measure for multipartite quantum systems. For system
of qubits we find that the difference between `classical' and `quantum'
regimes of correlation strength amounts to just 1 bit of information, at most.
We show that the information content of a correlation can be expanded into
correlations between pairwise components and demonstrate that in terms of this
information-based measure of correlation the GHZ states are the only states
that \textit{simultaneously} optimize these pairwise correlations for systems
of qubits.

\end{abstract}

\section{Introduction}

In classical communication theory it is clear that information cannot be
transmitted on a channel unless there is some \textit{correlation} between the
sent and received data [1]. If Alice and Bob, at the endpoints of the channel,
cannot establish such a correlation then they cannot exchange information. The
correlation is usually characterized by the mutual information $I\left(
X;Y\right)  $. The correlation between quantum systems is more interesting and
subtle than that between their classical counterparts, but this measure of
correlation has also been usefully applied to obtain insights into the
properties of correlated quantum systems [2-5].

With the discovery of discordant states of quantum systems it can be seen that
entanglement is not the only source of a correlation that is quantum
mechanical in nature [6,7]. The mutual information, appropriately defined,
between two quantum systems provides a measure of the total information
content contained in the correlation \ In particular, if $S$ is the total
entropy and $S\left(  k\right)  $ the entropy of a component part, then for a
2-component system the basis-independent single measure of correlation
strength, $S\left(  1\right)  +S\left(  2\right)  -S$, is not sensitive enough
on its own to distinguish the contributions to the correlation arising from
the various sources.

The relationship between entanglement and correlation and the `degree' of
non-classical behaviour does not seem to be wholly clear even for bipartite
entanglement. The difficulty is compounded when we look at tripartite
entanglements, or indeed multipartite entanglements. The entanglement of 3
quantum systems, for example, leads to the non-classical property of
\textit{monogamy} which places a limit on the ability to share entanglement
between the systems [8].

Accordingly here we focus simply on a measure of correlation \textit{strength}%
, largely avoiding the more difficult issues of the origin of that
correlation. Indeed, one may ask the question what does it mean to say that 2
(or more) quantum systems are correlated? In general, the correlation between
two systems is only made explicit when measurements are compared. That is to
say that correlation is exhibited in the joint properties of observables.

The spin-1/2 singlet state is an example of a maximally entangled bipartite
quantum state and yet there is no observable correlation between the spin-$z$
properties of one spin and the spin-$x$ properties of the other. A suitable
measure of overall correlation should, therefore, effectively `average' the
correlation between all possible choices of observable. The total information
content of the correlation between quantum systems, whether the contributions
to the correlation arise from classical or quantum properties, is a
fundamental metric that can be applied. As we shall see, this metric does
indeed distinguish between classical and non-classical regimes of correlation
\textit{strength}, but further analysis is necessary in order to elucidate the
particular non-classical properties that contribute to the stronger
correlations in the quantum regime.

Entanglement and discord, and their consequent effect on observable
correlations are quantum properties associated with the state of the system.
It is therefore reasonable to seek a measure of correlation that is
independent of any specific observable and captures some global property of
the state. Furthermore, we require that this measure be independent of any
notional splitting of a multicomponent system into various components. Of
course the correlations \textit{between} and \textit{within} any notional
partition can vary according to our particular choice, but the correlation of
the \textit{overall} system remains unaffected by any notional partitioning.

In this paper we examine a measure of correlation strength for multipartite
quantum systems, initially proposed in [4], that is simply the information
content of the correlation. This is defined as a natural extension of the
`index of correlation' for bipartite systems. The index of correlation for 2
systems is just the quantum generalization of the classical mutual information
and, for pure states, is equal to twice the entropy of entanglement. This
information-based measure arises as a consequence of imposing certain natural
conditions on any measure of correlation strength [9]. We show here that, in a
specific sense, this measure applied to systems of qubits implies that the
difference between `quantum' and `classical' is at most just 1 bit of
information, irrespective of the number of qubits. We examine the significance
of this single bit and show that the GHZ states of $n$-qubit systems are the
special states that simultaneously optimize pairwise correlations.

\section{Information as a Measure of Correlation}

\subsection{The Index of Correlation}

For 2 quantum systems the mutual information, defined by $I=S\left(  A\right)
+S\left(  B\right)  -S$, gives a measure of the strength of the correlation
between the systems $A$ and $B$. The entropies here are the quantum
generalizations of the classical entropy defined through the use of the
density operators as $S=~$Tr$\left\{  \rho\log\rho\right\}  $ and $S\left(
A\left[  B\right]  \right)  =~$Tr$\left\{  \rho_{A\left[  B\right]  }\log
\rho_{A\left[  B\right]  }\right\}  $. In previous work we termed this
measure, somewhat superfluously perhaps, as the \textit{index of correlation}.
This quantity is a direct measure of the information contained in the
correlation. The difference between quantum systems and classical systems can
be ascertained with use of the remarkable Araki-Lieb inequality [10] which
states that for any quantum systems $A$ and $B$ the entropies are related by%
\begin{equation}
\left\vert S\left(  A\right)  -S\left(  B\right)  \right\vert \leq S\leq
S\left(  A\right)  +S\left(  B\right)
\end{equation}
An immediate consequence of this is that for pure states of the $AB$ system
the sub-system entropies are equal. The LHS of this inequality is a
specifically quantum mechanical property and the equivalent inequality to (1)
for any two \textit{classical} systems is that%

\begin{equation}
\sup\left(  S\left(  A\right)  ,S\left(  B\right)  \right)  \leq S\leq
S\left(  A\right)  +S\left(  B\right)
\end{equation}
Using these properties we obtain the bounds for the strength of correlation
between two systems in the quantum and classical cases as%
\begin{align}
I_{Q}\left(  A;B\right)   &  \leq2\inf\left(  S\left(  A\right)  ,S\left(
B\right)  \right) \nonumber\\
& \nonumber\\
I_{C}\left(  A;B\right)   &  \leq\inf\left(  S\left(  A\right)  ,S\left(
B\right)  \right)
\end{align}
the bounds in each case only being reached when the sub-systems have equal
entropy, and in the quantum case when the total $AB$ system is also in a pure
state. If $\left\{  \left\vert a_{i}\right\rangle \right\}  $ and $\left\{
\left\vert b_{j}\right\rangle \right\}  $ are the eigenstates of the Schmidt
observables for the $A$ and $B$ systems, respectively, then assuming an
$n$-dimensional Hilbert space for both systems (for convenience) we have%
\[%
\begin{tabular}
[c]{|c|c|}\hline
State & Index of Correlation\\\hline
& \\
$\left\vert \psi_{AB}\right\rangle =%
%TCIMACRO{\dsum \limits_{j=1}^{n}}%
%BeginExpansion
{\displaystyle\sum\limits_{j=1}^{n}}
%EndExpansion
\sqrt{\pi_{j}}\left\vert a_{j},b_{j}\right\rangle $ & $-2%
%TCIMACRO{\dsum \limits_{j=1}^{n}}%
%BeginExpansion
{\displaystyle\sum\limits_{j=1}^{n}}
%EndExpansion
\pi_{j}\ln\pi_{j}$\\
& \\\hline
& \\
$%
%TCIMACRO{\dsum \limits_{j=1}^{n}}%
%BeginExpansion
{\displaystyle\sum\limits_{j=1}^{n}}
%EndExpansion
\pi_{j}\left\vert a_{j},b_{j}\right\rangle \left\langle a_{j},b_{j}\right\vert
$ & $-%
%TCIMACRO{\dsum \limits_{j=1}^{n}}%
%BeginExpansion
{\displaystyle\sum\limits_{j=1}^{n}}
%EndExpansion
\pi_{j}\ln\pi_{j}$\\
& \\\hline
& \\
$\rho_{A}\otimes\rho_{B}$ & 0\\
& \\\hline
\end{tabular}
\ \ \
\]
The region $\inf\left(  S\left(  A\right)  ,S\left(  B\right)  \right)  \leq
I\leq2\inf\left(  S\left(  A\right)  ,S\left(  B\right)  \right)  $ is clearly
a non-classical region. Only quantum states of our systems can possess a
degree of correlation strength in this region. However, we cannot make the
converse supposition that states in the region $0\leq I\leq$ $\inf\left(
S\left(  A\right)  ,S\left(  B\right)  \right)  $ are necessarily classical.
This is in accord with the profound result of Gisin and Peres which shows that
any entangled (non-factorable) pure state of 2 quantum systems will violate a
suitably-chosen Bell inequality [11].

The size of the index of correlation alone is thus not always sufficient to
determine whether a given state will display non-classical correlations. We
must compare this quantity with the individual sub-system entropies in order
to determine whether the state is non-classical. It is possible to find states
of 2 correlated systems which can be described by a separable mixed state
density operator with a higher value for the index of correlation than a
weakly-entangled pure state of the two systems.

A simple example of 2 correlated qubits illustrates this point; consider the
states of 2 qubits given by%
\begin{align}
\rho &  =\frac{1}{2}\left(  \left\vert 00\right\rangle \left\langle
00\right\vert +\left\vert 11\right\rangle \left\langle 11\right\vert \right)
\\
& \nonumber\\
\left\vert \psi\right\rangle  &  =a\left\vert 00\right\rangle +b\left\vert
11\right\rangle
\end{align}
The index of correlation for the first state is just $\ln2$ whereas the index
of correlation for the second can be lower than this when $\left\vert
a\right\vert \neq\left\vert b\right\vert $. The second state will, however,
violate a suitably-chosen Bell inequality [11] whereas the first state will
not, being essentially a classically-correlated state of the 2 qubits.
However, any state of $\left[  AB\right]  $ such that $\inf\left(  S^{\max
}\left(  A\right)  ,S^{\max}\left(  B\right)  \right)  \leq I\leq2\inf\left(
S^{\max}\left(  A\right)  ,S^{\max}\left(  B\right)  \right)  $ is definitely
a non-classical state that cannot be obtained by any classical description of
the 2 objects. For 2 qubits this means that any state of $\left[  AB\right]  $
such that $\ln2\leq I\left(  A;B\right)  \leq2\ln2$ is a non-classical state.

This simple example tells us that whilst the index of correlation is a good
measure of overall correlation strength, it is not on its own sufficient to
distinguish the nature of these correlations in all cases. It is only for
certain regions of correlation strength can we definitely say that the system
must be \textit{necessarily} non-classical. A state of the form (5) can be
very weakly entangled, with a very small value for any measure of `overall'
correlation, and yet still display highly non-classical properties for
specific observable correlations.

\subsection{The Index of Correlation for Multipartite Systems}

The total information content of the correlation between $N$ systems (each
possessing no degree of internal correlation) is the difference in the
information we obtain when considering the properties of each system on its
own and the information we obtain when considering the $N$-component system as
a whole [4]. The total information contained in the correlation is therefore
given by%

\begin{equation}
I\left(  1;2;\ldots;N\right)  =%
%TCIMACRO{\dsum \limits_{j=1}^{N}}%
%BeginExpansion
{\displaystyle\sum\limits_{j=1}^{N}}
%EndExpansion
S\left(  j\right)  -S
\end{equation}
where $S\left(  j\right)  $ is the entropy of the $j^{th}$ component
sub-system and $S$ is the total entropy. Consideration of this quantity tells
us that if we only measure the properties of the sub-systems of any
2-component quantum systems then we can only access at most half of the
information contained within the correlation [5]. The proof of this is based
on the Schmidt decomposition which is always possible for bipartite systems.
For tripartite, or multipartite, correlations it is not always possible to
obtain such a decomposition and so no such general and appealing result has
been so far obtained for $N>2$. However, application of (6) can still give us
useful insights into the correlation properties of $N$-component quantum systems.

The multipartite index of correlation (6) is a direct measure of the
information contained within all the correlations of an $N$-component system.
It is an observable-independent measure of correlation strength and as we
shall show it is invariant of any particular notional partitioning of our
quantum systems. Furthermore, it also satisfies the natural requirement that
any measure of correlation be \textit{additive} in the sense that $I\left(
\hat{\rho}\otimes\hat{\sigma}\right)  =I\left(  \hat{\rho}\right)  +I\left(
\hat{\sigma}\right)  $ for any 2 density operators $\hat{\rho}$ and
$\hat{\sigma}$, which of course implies additivity of the measure applied in
the same way to multipartite systems. It can be shown [9] that these 3
properties are sufficient to establish the index of correlation as the only
choice (up to the selection of logarithmic base).

\subsubsection{Internal and External Correlation}

Let us consider 4 systems each possessing no degree of internal correlation
which we label $A,B,C$ and $D$. We can (notionally) partition the 4-component
system into just 2 components so that, for example $\left[  AB\right]  $ is
considered to be a single component of the overall $ABCD$ system. We might
label the 2 components as $\alpha$ and $\beta$ so that for this example we
have $\alpha\equiv AB$ and $\beta\equiv CD$. From the definition of the index
of correlation (6) it is then easy to show that%
\begin{equation}
I\left(  ABCD\right)  =I\left(  AB\right)  +I\left(  CD\right)  +I\left(
\alpha,\beta\right)
\end{equation}
which we can rewrite in a different notation as%
\begin{equation}
I\left(  ABCD\right)  =I^{int}\left(  \alpha\right)  +I^{int}\left(
\beta\right)  +E\left(  \alpha,\beta\right)
\end{equation}
so that the overall index of correlation is simply the sum of the `internal
correlations' of $\alpha$ and $\beta$, denoted by $I^{int}\left(
\alpha\right)  $ and $I^{int}\left(  \beta\right)  ,$ and the `external
correlation' between the components $\alpha$ and $\beta$, which we have
denoted as $E\left(  \alpha,\beta\right)  $. This is an appealing result and
follows directly from the additivity properties of the index of correlation.

The overall correlation strength cannot depend on any notional split into
sub-components that we might choose and if we make a different choice of
partition denoted by a primed component then it can readily be shown that%
\begin{equation}
I^{int}\left(  \alpha\right)  +I^{int}\left(  \beta\right)  +E\left(
\alpha,\beta\right)  =I^{int}\left(  \alpha^{\prime}\right)  +I^{int}\left(
\beta^{\prime}\right)  +E\left(  \alpha^{\prime},\beta^{\prime}\right)
\end{equation}
In other words, as we change our perspective from the unprimed to the primed
components we change our perspective on the correlations \textit{within}
components and the correlations \textit{between} components. Our notion of
what is an internal and external correlation is \textit{relative} to our
choice of notional partition. In principle the choice of partition is
physically realizable by the simple expedient of spatially separating the
chosen components. We might envision, for example, a collection of $N=n+m$
qubits such that we give $n$ qubits to Alice and $m$ qubits to Bob.

It is not the purpose of this work to examine the measurement problem in
quantum mechanics, but we remark in passing that equation (9) may have
implications for this question. The division between `observed system' and
`measurement apparatus' in the Copenhagen interpretation is a matter of
judgement. The change in external correlation as we go from unprimed to primed
components can be thought of as a shifting of what we are including either
side of the system/apparatus boundary.

When a measurement is made, according to Copenhagen, there will be an abrupt
change in the external correlation $E\left(  \alpha,\beta\right)  $ between
system and apparatus. Before a measurement is made the change from unprimed to
primed is $\Delta E\left(  \alpha\rightarrow\alpha^{\prime}\right)  =-\Delta
I^{int}\left(  \alpha\rightarrow\alpha^{\prime}\right)  -\Delta I^{int}\left(
\beta\rightarrow\beta^{\prime}\right)  $ so that the change in external
correlation is the negative of the sum of the changes in internal correlation
as we change our perspective from unprimed to primed. If the object we are
calling our `apparatus' is to be thought of as a classical object in some
appropriate limit then as we continue to (notionally) call more of our
`apparatus' our `system' then at some point the roles of these will be reversed.

\subsubsection{Classical and Quantum Correlation Strengths}

One immediate and interesting consequence of (6) for $N$-qubit quantum systems
is the following. Consider a system of $N$ quantum sub-systems where, without
loss of generality, we have that%

\[
S\left(  1\right)  \geq S\left(  2\right)  \geq\ldots\geq S\left(  N\right)
\]
If these were classical systems we would have that%

\begin{equation}
\sup\left\{  S\left(  1\right)  ,S\left(  2\right)  ,\ldots,S\left(  N\right)
\right\}  \leq S\leq S\left(  1\right)  +S\left(  2\right)  +\ldots+S\left(
N\right)
\end{equation}
Hence, classically the upper bound on the total information content of the
correlation is given by%

\begin{equation}
I_{C}\left(  1;2;\ldots;N\right)  \leq%
%TCIMACRO{\dsum \limits_{j=2}^{N}}%
%BeginExpansion
{\displaystyle\sum\limits_{j=2}^{N}}
%EndExpansion
S\left(  j\right)
\end{equation}
and we have used the subscript $C$ to denote classical. For quantum systems,
however, the total entropy can be equal to zero so that an upper bound for the
information content of the correlation is%

\begin{equation}
I_{Q}\left(  1;2;\ldots;N\right)  \leq%
%TCIMACRO{\dsum \limits_{j=1}^{N}}%
%BeginExpansion
{\displaystyle\sum\limits_{j=1}^{N}}
%EndExpansion
S\left(  j\right)
\end{equation}
The difference between the two quantities is bounded by%

\begin{equation}
I_{Q}\left(  1;2;\ldots;N\right)  -I_{C}\left(  1;2;\ldots;N\right)  \leq
S\left(  1\right)
\end{equation}
For quantum systems of $N$ qubits, therefore, we have the following result

\begin{quote}
the difference between the maximal correlation of $N$ qubits and the maximum
correlation that can be achieved by a classical system of $N$ bits is just 1
bit, \textit{independent of} $N$
\end{quote}

However many qubits we entangle together the extra information we have in the
correlations in the quantum regime is at most 1 bit. For qubits, then, the
difference between classical and quantum, in this sense, amounts to just 1 bit
of extra information. In other words, as we increase $N$ we can see that the
relative gain in using entangled quantum qubits decreases as $N^{-1}$. This is
an interesting consequence of (6) because we might suppose that increasing the
`available entanglement' by entangling more and more qubits together would
lead to a corresponding increase in capability over any corresponding
classical equivalent system. Another way of stating this result is that any
extra capability, due to correlation, arising from entangling quantum qubits
together arises from just 1 extra bit of information in the correlations.

In a loose sense, for collections of many qubits it might be generally
supposed that the `amount' of entanglement is directly related to the amount
of quantum `magic' that can be performed. Of course this term is somehat
whimsical but the counter-intuitive result that the quantum regime of
correlation \textit{strength} amounts to just 1 extra bit indicates that this
single measure of correlation strength is not sensitive, or subtle, enough to
distinguish all of the potential non-classical features. It is an open
question at the moment as to the precise nature of the relationship between
correlation strength, as given by the index of correlation, and what might be
termed non-classical behaviour (or quantum `magic' if we wish to be more whimsical).

There are many measures of `non-classical' and it is not currently clear to me
what the relationship is between these measures and the non-classical region
outlined by the index of correlation. For example, for bosonic systems the
Glauber-Sudarshan representation affords a precise measure of non-classical.
The index of correlation, as discussed above, is only providing a measure of
non-classical for a particular region, namely those states for which the
correlation strength lies between the classical and quantum upper bounds.
Further work is necessary to fully elucidate this relationship.

\section{Systems of Multiple Qubits}

The first situation where we might expect to find significant differences
between bipartite correlations and multipartite correlations is when $N=3$ and
we have tripartite entanglement. We shall label our 3 systems as $A,B,$ and
$C$ and, where necessary, assume as before that $S\left(  A\right)  \geq
S\left(  B\right)  \geq S\left(  C\right)  $. It is tempting to consider the
inclusion-exclusion principle for sets as giving us an adequate description of
the correlation between 3 systems. Classically, however, the interpretation of
such Venn-diagrams for entropies can be misleading since there are regions
which can be negative [12]. In terms of entropies this use of the
inclusion-exclusion principle leads to a parameter $\Lambda$ given by%

\begin{equation}
\Lambda=S\left(  A\right)  +S\left(  B\right)  +S\left(  C\right)  -\left\{
S\left(  AB\right)  +S\left(  AC\right)  +S\left(  BC\right)  \right\}  +S
\end{equation}
It is easy to see that for pure states of 3 quantum systems then $\Lambda=0$,
whatever their state of correlation. We also have that $\Lambda=0$ for
uncorrelated systems.

The total information content of the correlation between three systems is
given from (6) by%

\begin{equation}
I\left(  \left[  ABC\right]  \right)  =S\left(  A\right)  +S\left(  B\right)
+S\left(  C\right)  -S
\end{equation}
where we adopt the $\left[  \ldots\right]  $ notation to indicate that the
quantities inside the bracket are to be taken as a single system. Using this
notation we see that $I\left(  A;\left[  BC\right]  \right)  $ denotes the
information content of the correlation between sub-system $A$ and sub-systems
$\left[  BC\right]  $, taken as a single sub-system. Equation (15) can be
written in the form%

\begin{equation}
I\left(  \left[  ABC\right]  \right)  =I\left(  A;\left[  BC\right]  \right)
+I\left(  B;C\right)
\end{equation}
where $I\left(  A;\left[  BC\right]  \right)  =S\left(  A\right)  +S\left(
BC\right)  -S$. If we think of $A$ as our `external' sub-system, then the
total tripartite correlation is made up of the external correlation between
$A$ and $\left[  BC\right]  $ and the `internal' correlation that exists
between sub-systems $B$ and $C$. It should be noted that $I\left(  \left[
XY\right]  \right)  =I\left(  X;Y\right)  $ \textit{only if} the partitions
$X$ and $Y$ contain no internal correlation.

\subsection{Correlation Interpretation of Strong Subadditivity}

The use of an information-based metric for correlation leads to the following
well-known fundamental result. Let us consider the external correlation
between between $A$ and $\left[  BC\right]  $ which is given by $I\left(
A;\left[  BC\right]  \right)  $. Clearly, there cannot be less information in
the correlation between $A$ and $\left[  BC\right]  $ than there is between
either $A$ and $B$ alone. In other words, we have that%
\begin{equation}
I\left(  A;\left[  BC\right]  \right)  \geq I\left(  A;B\right)
\end{equation}
Hence%
\begin{equation}
S\left(  A\right)  +S\left(  BC\right)  -S\geq S\left(  A\right)  +S\left(
B\right)  -S\left(  AB\right)
\end{equation}
which, upon rearrangement, gives us the strong subaddivity relation for
tripartite systems%
\begin{equation}
S\left(  AB\right)  +S\left(  BC\right)  \geq S\left(  B\right)  +S
\end{equation}
Strong subadditivity can, therefore, be seen as arising from the natural
condition on the information content of the correlations between the component
sub-systems expressed by (17).

If we rewrite the strong subadditivity relation $S\left(  AC\right)  +S\left(
BC\right)  \geq S\left(  C\right)  +S$ in terms of the correlation information
we have that%
\begin{equation}
I\left(  \left[  ABC\right]  \right)  \geq I\left(  A;C\right)  +I\left(
B;C\right)
\end{equation}
Strong subadditivity therefore implies that the total correlation of any
3-component quantum system is greater than (or equal to) the sum of any two of
the correlation of the pairwise correlations $I\left(  A;B\right)  ,I\left(
A;C\right)  $ and $I\left(  B;C\right)  $.

\subsection{Pure States of 3-Component Systems}

For \textit{pure} states of the 3-component system we have that%
\begin{equation}
I^{pure}\left(  \left[  ABC\right]  \right)  =I\left(  A;B\right)  +I\left(
A;C\right)  +I\left(  B;C\right)
\end{equation}
which together with (20) suggests the bounds%
\begin{equation}
I\left(  A;C\right)  +I\left(  B;C\right)  \leq I\left(  \left[  ABC\right]
\right)  \leq I\left(  A;B\right)  +I\left(  A;C\right)  +I\left(  B;C\right)
\end{equation}
It is worth noting here that the GHZ states are the unique pure states of 3
qubits that maximise the index of correlation $I\left(  \left[  ABC\right]
\right)  $ [13].

The difference between the information content of the correlation between the
external system and the components of the sub-system $\left[  BC\right]  $ and
the information content of the external correlation $I\left(  A;\left[
BC\right]  \right)  $ is therefore giving us a measure of the `wholeness' of
the sub-sytem $\left[  BC\right]  $. If the properties of $B$ are wholly
determined by knowledge of the properties of $C$ then we would expect that
$I\left(  A;\left[  BC\right]  \right)  =I\left(  A;B\right)  =I\left(
A;C\right)  $. The difference is given by%

\begin{equation}
I\left(  A;B\right)  +I\left(  A;C\right)  -I\left(  A;\left[  BC\right]
\right)  =\Lambda
\end{equation}
where $\Lambda$ is the set-counting parameter given by equation (14). For pure
states of the total 3-component system, therefore, we have the remarkable
relation that%

\begin{equation}
I\left(  A;B\right)  +I\left(  A;C\right)  =I\left(  A;\left[  BC\right]
\right)
\end{equation}
which tells us that, for pure states of $\left[  ABC\right]  $ the external
correlation is simply the sum of the correlation between the external system
$A$ and the individual components of the sub-system $\left[  BC\right]  $.

\subsubsection{Entropy Bounds on 3-Component Pure States}

The most important difference between classical and quantum systems, from the
perspective of correlation, is that correlated quantum systems can be prepared
in pure states. The Araki-Lieb inequality (1) for a 2-component partition
highlights this in a dramatic way.

That bipartite quantum systems have a lower entropy bound than their classical
counterparts is a remarkable property; the smaller component of the total
system placing a quite severe limit on the allowable uncertainty of the larger
component when the total $\left[  AB\right]  $ system is in a pure state. If,
for example, $A$ is some field state and $B$ a qubit, such as a two-level
atom, then the uncertainty in the field is, at most, 1 bit when the atom-field
system is in a pure state. The field can, in this situation, be fully
described by just 2 states [14]. 

A \textit{necessary}, but not \textit{sufficient}, condition for states of
$N$-component quantum systems to maximise the total information content in the
correlation is that the total system is in a pure state. The possibility of
preparing correlated quantum systems in pure states gives quantum systems the
possibility of containing more information in their correlation than can be
achieved if those systems were to be classical. Thus, if we were to attempt to
model the previous simple example of a single 2-level atom interacting with a
field using a classical or semi-classical treatment we would find that this
classical description would not be able to reproduce the correlation strength
available for some of the initial states afforded by the quantum description.

For pure states of 3-component quantum systems we find that the smallest
component also places a quite severe limit on the allowable entropies of the
remaining sub-systems as follows. As we have seen, strong subadditivity
follows from the condition on the correlations that%
\begin{equation}
I\left(  C;\left[  AB\right]  \right)  \geq\sup\left\{  I\left(  A;C\right)
,I\left(  B;C\right)  \right\}
\end{equation}
For pure states of the $\left[  ABC\right]  $ system we have, from the
Araki-Lieb inequality (1) that $S\left(  AB\right)  =S\left(  C\right)  $, and
cyclic permutations, so that%
\begin{equation}
S\left(  C\right)  \geq\left\vert S\left(  A\right)  -S\left(  B\right)
\right\vert
\end{equation}
If we assume that $C$ is the component with the smallest dimensionality, then
this smallest component places a restriction on the \textit{difference}
between the entropies of the remaining sub-systems, whatever their size. If,
for example, $C$ is a qubit, then the difference between the entropies of the
$A$ and $B$ systems, whatever their size, can be at most 1 bit for pure states
of the $\left[  ABC\right]  $ system.

Equation (26) is simply an expression of the requirement that the infomation
content of the correlation between any two systems must be greater than or
equal to zero so that, for example, $I\left(  B;C\right)  \geq0$. This
requirement also implies strong sub-additivity for pure states. To see this we
note that for pure states of $\left[  ABC\right]  $ we have that $I\left(
A;B\right)  +I\left(  A;C\right)  =2S\left(  A\right)  $. The condition
$I\left(  B;C\right)  \geq0$ which gives $S\left(  A\right)  \leq S\left(
B\right)  +S\left(  C\right)  $ then implies%

\begin{equation}
I\left(  A;B\right)  +I\left(  A;C\right)  \leq S\left(  A\right)  +S\left(
B\right)  +S\left(  C\right)
\end{equation}
from which strong subadditivity for pure states follows. Strong subadditivity
for \textit{pure} states is, therefore, nothing more than the requirement that
the pairwise mutual information of the sub-systems be non-negative.

\section{Partitions and Maximal Correlation of Qubits}

Let us consider a general quantum system that is comprised of $n$ component
sub-systems so that $H=H_{1}\otimes H_{2}\otimes\ldots\otimes H_{n}$. We shall
assume these component sub-systems possess no degree of internal correlation.
In general we can assign an arbitrary partition on the space with the number
of distinct assignable partitions being approximated by the asymptotic
expression%
\begin{equation}
p\left(  n\right)  \sim\frac{1}{4n\sqrt{3}}\exp\left(  \pi\sqrt{\frac{2n}{3}%
}\right)  ~~~~~~\text{as }n\rightarrow\infty
\end{equation}
As an example let us consider the partitioning of a general quantum system
into 3 parts $A,B$ and $C$ with the number of component sub-systems in each
part being given by $a,b$ and $c$, respectively, where $a+b+c=n$. The total
information content of the entire $n$-component system can be expressed as%

\begin{equation}
I\left(  \left[  1;2;3;\ldots;n\right]  \right)  =I\left(  \left[  A\right]
;\left[  B\right]  ;\left[  C\right]  \right)  +I\left(  \left[  A\right]
\right)  +I\left(  \left[  B\right]  \right)  +I\left(  \left[  C\right]
\right)
\end{equation}
where $I\left(  \left[  A\right]  \right)  $ represents the `internal'
correlation of the part $A$, for example. If any of our $A,B$ and $C$ are the
fundamental component systems (assumed to possess no degree of internal
correlation) then we can set internal correlation contribution to zero.

Let us consider the partitioning of the space into just two parts which we
shall label $P$ and $Q$ so that
\begin{align}
H  &  =\left(  H_{1}\otimes H_{2}\otimes\ldots H_{p}\right)  \otimes\left(
H_{p+1}\otimes H_{p+2}\otimes\ldots H_{p+q}\right) \nonumber\\
&  =H_{P}\otimes H_{Q}%
\end{align}
If there exists some measure of correlation, $\gamma$,\ of a state then there
will be some state in the space which maximises this quantity. In general this
maximum value which we denote by $\tilde{\gamma}$, will depend on the
dimensionality of the space. If $P$ and $Q$ are independent (uncorrelated)
sub-spaces of $H$ then there exists a state $\left\vert \xi\right\rangle
=\left\vert \mu_{P}\right\rangle \otimes\left\vert \mu_{Q}\right\rangle $ in
$H$ where $\left\vert \mu_{P}\right\rangle $ and $\left\vert \mu
_{Q}\right\rangle $ are the states that maximise this measure of correlation
in $P$ and $Q$ respectively. The maximum correlation that can be attained by a
state in $H$ must therefore be such that%
\begin{equation}
\tilde{\gamma}_{H}\left(  \dim H\right)  \geq\tilde{\gamma}_{P}\left(  \dim
H_{P}\right)  +\tilde{\gamma}_{Q}\left(  \dim H_{Q}\right)
\end{equation}
For the index of correlation we have that $I\left(  H\right)  \leq I\left(
P\right)  +I\left(  Q\right)  $ and so the maximum values must satisfy%
\begin{equation}
\tilde{I}\left(  H\right)  =\tilde{I}\left(  P\right)  +\tilde{I}\left(
Q\right)
\end{equation}
where the tilde denotes the maximum value. For $n\geq4$ we can always
partition the qubits such that the maximum possible value for $I$ over all the
qubits and states is the same as we would obtain by considering the parts of
this partition to be uncorrelated and separately maximising the index of
correlation for the individual parts.

So, for example, if we have 6 qubits then the maximum possible value for $I$
over all states is 6 bits. We can however partition this space into 2 spaces
each containing 3 qubits. The state $\left\vert \psi\right\rangle =\left\vert
\psi_{GHZ}\right\rangle \otimes\left\vert \psi_{GHZ}\right\rangle $ then
achieves the maximum possible value for $I$ even though this state represents
a separable system. The partition $\left(  2,2,2\right)  $ in which each
2-qubit part is in a Bell state also gives the maximum possible value for $I$
of 6 bits.

The 2 and 3 qubit systems are `irreducible' in this way; there is no partition
we can make such that separately maximising the correlation in the parts gives
the maximum correlation for the whole space. For these systems we need the
Bell or GHZ states to achieve the maximal correlation (the Bell states can, of
course, be viewed as the GHZ states for 2 qubits). For $n\geq4$ we can always
partition the space into parts containing only 2 or 3 qubits. Indeed, we can
always find integers $p$ and $q$ such that $n=2p+3q$. Clearly, there will, in
general, be many ways we can partition the space of a large number of qubits
into these irreducible parts.

\subsection{The Special Role of GHZ States}

Let us consider $n$ systems, which for convenience we shall take to be qubits,
although the treatment is, of course, applicable to more general systems. For
$n=2$ there are two integer partitions we can make which we can describe by
the sequences (2) and (1,1). In terms of the index of correlation these two
partitions are equivalent. For $n=3$ there are 3 partitions which we can
describe by the sequences (3), (2,1) and (1,1,1) and again, the (3) and
(1,1,1) partitions are equivalent as far as the index of correlation is
concerned. We can think of each partition as corresponding to a particular
choice of expansion for the index of correlation.

The partitioning and subsequent identification of internal and external
correlations allows us to recursively treat the total information content of
the correlations between $n$ qubits so that if we label our qubits as
$1,2,3,\ldots,n$ we can write%
\begin{equation}
I\left(  \left[  1\ldots n\right]  \right)  =I\left(  \left[  2\ldots
n\right]  \right)  +I\left(  1;\left[  2\ldots n\right]  \right)
\end{equation}
but $I\left(  \left[  2\ldots n\right]  \right)  $ is just the correlational
information content of an $n-1$ qubit system (whose state is constrained by
the given $n$-qubit state). Clearly this kind of recursive procedure for the
index of correlation can be applied more generally to a total system of
correlated (or interacting) quantum sub-systems where the sub-systems do not
have to be qubits (or even identical). Applying this recursion we obtain the
expansion of $I\left(  \left[  1\ldots n\right]  \right)  $ into $n-1$
pairwise correlations between a single qubit and the remaining qubits as%
\begin{equation}
I\left(  \left[  1\ldots n\right]  \right)  =%
%TCIMACRO{\dsum \limits_{k=1}^{n-1}}%
%BeginExpansion
{\displaystyle\sum\limits_{k=1}^{n-1}}
%EndExpansion
I\left(  k;\left[  \left(  k+1)\ldots\left(  n-1\right)  n\right)  \right]
\right)
\end{equation}
Thus for $n=4$ we have that%
\begin{equation}
I\left(  \left[  1234\right]  \right)  =I\left(  1;\left[  234\right]
\right)  +I\left(  2;\left[  34\right]  \right)  +I\left(  3;4\right)
\end{equation}
Equation (34) can be thought of in two ways. Either we can consider the whole
system and successively `remove' a qubit, or we can consider just 2 qubits and
successively `add' a qubit until we have our $n$ qubits.

Let us now consider the states that maximise $I\left(  \left[  1\ldots
n\right]  \right)  $. Clearly, we require a pure state of the $\left[  1\ldots
n\right]  $ system. As we have seen, partitions of the space into 2 and 3
qubits in which these parts are not correlated with one another can give us a
variety of states for which the total correlation $I\left(  \left[  1\ldots
n\right]  \right)  $ is maximised. A tensor product state of Bell and
3-component GHZ states will maximise this quantity. However, we can ask what
\textit{non-separable} states of $\left[  1\ldots n\right]  $ will maximise
the correlation? Obviously the GHZ state%
\begin{equation}
\left\vert \psi\right\rangle =\frac{1}{\sqrt{2}}\left(  \left\vert
0\right\rangle ^{n}+\left\vert 1\right\rangle ^{n}\right)
\end{equation}
is an example of such a state. We now show that the states of the GHZ form are
rather special in that they are the only states that \textit{simultaneously}
optimise the pairwise correlations occurring in the expansion (34).

We begin with just 2 qubits, the $n^{th}$ and the $\left(  n-1\right)  ^{th}$
qubits, and consider the correlation $I\left(  n-1;n\right)  $. We are seeking
non-separable states that maximise $I\left(  \left[  1\ldots n\right]
\right)  $ and so the $\left[  \left(  n-1\right)  n\right]  $ state cannot be
a pure state part of the partition. The quantum bound on the correlation
between 2 systems gives us $I\left(  n-1;n\right)  \leq2\inf\left(  S\left(
n-1\right)  ,S\left(  n\right)  \right)  $ but if we achieve close to the
upper bound of 2 bits then this implies the $n-2$ qubit is almost decoupled
from the $\left[  \left(  n-1\right)  n\right]  $ part of the partition and we
have that $I\left(  n-2;\left[  \left(  n-1\right)  n\right]  \right)
\approx0$ so that achieving close to the maximal degree of correlation for 2
qubits gives us an almost minimal degree of correlation between these 2 qubits
and any other qubit and so the pairwise correlations are not simultaneously optimised.

Furthermore, if we consider $I\left(  1;\left[  2\ldots n\right]  \right)  $
which is the correlation between the last qubit we `add' and the rest of the
qubits we see that its maximal value can be 2 bits since the overall state is
pure. The information content of the correlation the remaining $\left[
2\ldots n\right]  $ qubits can therefore be at most $n-2$ bits. This implies
that the information content of each pairwise correlation $I\left(  k;\left[
k\left(  k+1)\ldots\left(  n-1\right)  n\right)  \right]  \right)  $ must be 1
bit for $k\geq2$ if we wish to simultaneously optimise all of the pairwise correlations.

If the correlation between the $n^{th}$ and the $\left(  n-1\right)  ^{th}$
qubits is 1 bit this implies that the entropies $S\left(  \left[  \left(
n-1\right)  n\right]  \right)  =S\left(  n-1\right)  =S\left(  n\right)  =1$
bit. The state of these qubits must then be described by a density operator of
the form%
\begin{equation}
\rho_{\left[  \left(  n-1\right)  n\right]  }=\frac{1}{2}\left(  \left\vert
ab\right\rangle \left\langle ab\right\vert +\left\vert \bar{a}\bar
{b}\right\rangle \left\langle \bar{a}\bar{b}\right\vert \right)
\end{equation}
where $a,b\in\left\{  0,1\right\}  $ and the bar denotes the bit complement.
The states $\left\vert ab\right\rangle =\left\vert a\right\rangle
\otimes\left\vert b\right\rangle $ refer to some suitable basis for the
individual qubits which could represent different spin directions for each of
the 2 qubits. Without any significant loss of generality we shall take the
basis to be the computational basis (spin-$z$ for both qubits) and set $a=b=0$.

Let us now add in the next qubit, the $\left(  n-2\right)  ^{th}$ qubit. The
pairwise correlation between this and our original 2 qubits is $I\left(
n-2;\left[  \left(  n-1\right)  n\right]  \right)  $ which is just 1 bit. The
density operator for the 3 qubits must therefore be of the form%
\begin{align}
\rho_{\left[  \left(  n-2\right)  \left(  n-1\right)  n\right]  }  &
=\frac{1}{2}\left\vert 0\right\rangle \otimes\left\vert \varphi_{\left[
\left(  n-1\right)  n\right]  }\right\rangle \left\langle \varphi_{\left[
\left(  n-1\right)  n\right]  }\right\vert \otimes\left\langle 0\right\vert
\nonumber\\
&  +\frac{1}{2}\left\vert 1\right\rangle \otimes\left\vert \chi_{\left[
\left(  n-1\right)  n\right]  }\right\rangle \left\langle \chi_{\left[
\left(  n-1\right)  n\right]  }\right\vert \otimes\left\langle 1\right\vert
\end{align}
where we have again assumed the computational basis for the $\left(
n-2\right)  ^{th}$ qubit. Comparison of the reduced density operator
$\rho_{\left[  \left(  n-1\right)  n\right]  }$ obtained from (67) and that of
(66) gives us that%
\begin{align}
\left\vert \varphi_{\left[  \left(  n-1\right)  n\right]  }\right\rangle  &
=\left\vert 00\right\rangle \nonumber\\
\left\vert \chi_{\left[  \left(  n-1\right)  n\right]  }\right\rangle  &
=\left\vert 11\right\rangle
\end{align}
the particular form being dictated by our choice of the computational basis
and the bit values. In general, we would obtain $\left\vert \varphi_{\left[
\left(  n-1\right)  n\right]  }\right\rangle =\left\vert ab\right\rangle $
with $\left\vert \chi_{\left[  \left(  n-1\right)  n\right]  }\right\rangle
=\left\vert \bar{a}\bar{b}\right\rangle $ where the individual states in the
tensor product are expressed in some appropriate spin basis. Continuing to add
qubits in this fashion yields the density operator for the $\left[  2\ldots
n\right]  $ part of the partition%
\begin{equation}
\rho_{\left[  2\ldots n\right]  }=\frac{1}{2}\left(  \left\vert 00\ldots
0\right\rangle \left\langle 00\ldots0\right\vert +\left\vert 11\ldots
1\right\rangle \left\langle 11\ldots1\right\vert \right)
\end{equation}
Adding in the final qubit is equaivalent to a purification of this density
operator and we obtain the final pure state for the total $\left[  1\ldots
n\right]  $ system of $n$ qubits as the GHZ state%
\begin{equation}
\left\vert \psi\right\rangle =\frac{1}{\sqrt{2}}\left(  \left\vert
0\right\rangle ^{n}+\left\vert 1\right\rangle ^{n}\right)
\end{equation}
In general, therefore, the state that \textit{simultaneously} optimises the
pairwise correlations of the total $\left[  1\ldots n\right]  $ system of $n$
qubits is given by%
\begin{equation}
\left\vert \psi\right\rangle =\frac{1}{\sqrt{2}}\left(  \left\vert b_{1}%
b_{2}\ldots b_{n}\right\rangle +\left\vert \bar{b}_{1}\bar{b}_{2}\ldots\bar
{b}_{n}\right\rangle \right)
\end{equation}
where $b_{k}\in\left\{  0,1\right\}  $ and the bar denotes the bit complement,
as before. The states that simultaneously optimise the pairwise correlations
are therefore of the GHZ form.

We now see why, in terms of correlation strength, the difference between
quantum and classical for a system of $n$ qubits is at most 1 bit. At each
stage in the addition of qubits to our original 2 we have the most correlated
classical state for the sub-system. It is only when we get to the last qubit
that the quantum nature becomes manifest when we purify the $\rho_{\left[
2\ldots n\right]  }$ density operator.

\section{Conclusion}

The production and manipulation of correlated systems of qubits where the
quantum nature of the correlation can be used as a resource to yield
properties unachievable within a classical framework is a very active and
important area of research. The potential inherent in exploiting entanglement
can be seen in the development of such new and exciting technologies such as
quantum teleportation [15] and quantum computation [16]. Indeed the whole area
of quantum information processing in general seems to be a rich framework for
the development of new capabilities and insights [17] and the notions of
entanglement have generated new perspectives in areas as diverse as, biology
[18], thermodynamics [19] and games [20], to give just 3 examples of many.

It would seem therefore that understanding the nature of the correlation
between quantum systems is an important goal. Whilst correlated states of 2
systems, such as the celebrated singlet state for spin-1/2 particles, may seem
simple and straightforward on the surface, at least from a purely mathematical
perspective, even pairwise quantum correlations yield a surprisingly subtle
and difficult behaviour to interpret. The introduction of the idea of discord
has beautifully emphasized that there is still much to learn even about such
simple correlated systems. The difficulties and subtleties are further
compounded when we consider correlations between more than just 2 quantum
systems. The property of monogamy elegantly illustrates some of the further
subtleties encountered when we move beyond bipartite correlations.

Information is a fundamental metric that finds a natural role in the
description of correlation. Here we have applied this parameter purely as a
measure of correlation \textit{strength} without concerning ourselves about
the particular quantum or classical feature that is contributing to the
correlation. From the perspective afforded by this information-based parameter
we have argued that the difference between `classical' and `quantum' for
systems of qubits amounts to, at most, one additional bit of information
contained in the correlation. We have further shown that for tripartite pure
states the entropy of one sub-system places a fundamental bound on the
\textit{difference} in the entropies of the remaining 2 sub-systems.

By developing an expansion of multi-qubit correlation into pairwise
correlations of single qubits and collections of qubits it is possible to view
the GHZ states as a fundamental state of systems of qubits in that states of
this form are the only ones that simultaneously optimise these pairwise
correlations. The optimization here is a constrained simultaneous optimization
of the various components because, as the property of monogamy demonstrates,
the individual pairwise correlations are not independent quantities. \bigskip

\textbf{Acknowledgements\bigskip}

I would like to thank S.M. Barnett, F.S. Khan and N. L\={u}tkenhaus, for
valuable and enlightening discussions.\bigskip

\section{References}

\begin{enumerate}
\item See, for example, R. W. Hamming, \textit{Coding and Information Theory},
Prentice-Hall, New Jersey, $\left(  1980\right)  $

\item W. H. Zurek, \textit{Pointer Basis of Quantum Apparatus: Into what
Mixture does the Wave Packet Collapse}?, Phys. Rev D, \textbf{24}, 1516--1525,
$(1981)$; W. H. Zurek, \textit{Environment-Induced Superselection Rules},
Phys. Rev. D, \textbf{26}, 1862--1880, $(1982)$; W. H. Zurek,
\textit{Decoherence, einselection, and the quantum origins of the classical},
Rev. Mod. Phys., \textbf{75}, 715, $(2003)$

\item S. M. Barnett and S. J. D. Phoenix,\textit{\ Information Theory,
Squeezing and Quantum Correlations}, Phys. Rev. A, \textbf{44}, 535, $(1991)$

\item S. M. Barnett and S. J. D. Phoenix, \textit{Information-Theoretic Limits
to Quantum Cryptography}, Phys. Rev. A, \textbf{48,} R5, $(1993)$

\item S. M. Barnett and S. J. D. Phoenix, \textit{Bell's Inequality and the
Schmidt Decomposition}, Phys. Lett. A, \textbf{167,} 233, $(1992)$

\item H. Ollivier and W. H. Zurek, \textit{Quantum Discord: A Measure of the
Quantumness of Correlations}, Phys. Rev. Lett. \textbf{88}, 017901, $(2001)$;
L. Henderson and V. Vedral, \textit{Classical, Quantum and Total
Correlations}, J. Phys. A \textbf{34}, 6899, $(2001)$

\item B. Daki\'{c}, V. Vedral, and C. Brukner, \textit{Necessary and
Sufficient Condition for Nonzero Quantum Discord}, Phys. Rev. Lett.,
\textbf{105}, (19), 190502, $(2010)$

\item V. Coffman, J. Kundu and W. K. Wootters, \textit{Distributed
Entanglement}, Phys. Rev. A \textbf{61}, 052306, $(2000)$

\item S.J.D .Phoenix and F.S. Khan, \textit{in preparation}

\item H. Araki and E. H. Lieb, \textit{Entropy Inequalities}, Comm. Math.
Phys., \textbf{18}, 160, $(1970)$

\item N. Gisin and A. Peres, \textit{Maximal Violation of Bell's Inequality
for Arbitrarily Large Spin}, Phys. Lett. A \textbf{162}, 15 $(1992)$

\item D.J.C. MacKay, \textit{Information Theory, Inference, and Learning
Algorithms}, CUP, Cambridge, $(2003)$

\item S.M. Barnett, \textit{private communication}

\item S.J.D. Phoenix and P.L. Knight, \textit{Fluctuations and Entropy in
Models of Optical Resonance}, Ann. Phys. (NY), \textbf{186,} 381, $(1988)$

\item C. H. Bennett, G. Brassard, C. Cr\'{e}peau, R. Jozsa, A. Peres, and W.
K. Wootters, \textit{Teleporting an Unknown Quantum State via Dual Classical
and Einstein-Podolsky-Rosen Channels}, Phys. Rev. Lett. \textbf{70}, 1895,
$(1993)$; F. Bussieres, C. Clausen, A. Tiranov, B. Korzh, V. B. Verma, S. W.
Nam, F. Marsili, A. Ferrier, P. Goldner, H. Herrmann, C. Silberhorn, W.
Sohler, M. Afzelius, and N. Gisin, \textit{Quantum Teleportation from a
Telecom-Wavelength Photon to a Solid-State Quantum Memory},
http://arxiv.org/abs/1401.6958 $\left(  2014\right)  $

\item P. W. Shor, \textit{Polynomial-Time Algorithms for Prime Factorization
and Discrete Logarithms on a Quantum Computer}, SIAM J. Comput., \textbf{26}
(5), 1484, $(1997)$

\item S.M. Barnett, \textit{Quantum Information}; Oxford University Press:
Oxford, $\left(  2009\right)  $

\item M. Sarovar, A\ . Ishizaki, G.R. Fleming, K. B. Whaley, \textit{Quantum
Entanglement in Photosynthetic Light-Harvesting Complexes},. Nature Physics
\textbf{6} (6): 462, $(2010)$

\item J. Ro\ss nagel, O. Abah, F. Schmidt-Kaler, K. Singer and E. Lutz,
\textit{Nanoscale Heat Engine Beyond the Carnot Limit},
http://arxiv.org/pdf/1308.5935v2, $\left(  2014\right)  $

\item J. Eisert, M. Wilkens and M. Lewenstein, \textit{Quantum Games and
Quantum Strategies}, Phys. Rev. Lett., \textbf{83}, 3077-3080, $(1999)$; D.
Meyer, \textit{Quantum Strategies, }Phys. Rev. Lett., \textbf{82}, 1052-1055,
$(1999)$; A. Iqbal and S. Weigert, \textit{Quantum Correlation Games}, J.
Phys. A, \textbf{37}, 5873-5885, $(2004)$; C.D. Hill, A.P. Flitney and N.C.
Menicucci, \textit{A Competitive Game whose Maximal Nash-equilibrium Payoff
Requires Quantum Resources for its Achievement}, Phys. Lett. A \textbf{374,
}3619 $(2010)$; J. Shimamura, S. \"{O}zdemir, F. Morikoshi, and N. Imoto,
\textit{Quantum and Classical Correlations Between Players in Game Theory},
Int. J. Quant. Inf., \textbf{2}, 79--89 $(2004)$; S.J.D .Phoenix and F.S.
Khan, \textit{The Role of Correlation in Quantum and Classical Games},
Fluctuation and Noise Letters, \textbf{12}, 1350011, $(2013)$
\end{enumerate}

\end{document}